\documentclass[conference]{IEEEtran}
\IEEEoverridecommandlockouts
\usepackage[utf8]{inputenc}
\usepackage[T1]{fontenc}
\usepackage{textcomp}
\usepackage{cite}
\usepackage{amsmath,amssymb,amsfonts}
\usepackage{algorithmic}
\usepackage{graphicx,subcaption}
\usepackage{xcolor}
\usepackage{soul}
\usepackage{float}
\usepackage{stfloats}
\usepackage{pgfplots}
\pgfplotsset{compat=1.17}
\usepgfplotslibrary{statistics}
\usepackage[shortcuts,acronym]{glossaries}
\usepackage{booktabs}
\usepackage{balance}
\usepackage[bookmarks=false]{hyperref}

\def\BibTeX{{\rm B\kern-.05em{\sc i\kern-.025em b}\kern-.08em
    T\kern-.1667em\lower.7ex\hbox{E}\kern-.125emX}}
\begin{document}
\title{
Rethinking NB-IoT Downlink Synchronization for LEO-NTN: A Novel Overhead Reduction Method and Measurement-Based Evaluation
}

\author{\IEEEauthorblockN{\"{O}mer L\"utf\"u Karakelle\IEEEauthorrefmark{1}\IEEEauthorrefmark{2}, Erhan Karakoca\IEEEauthorrefmark{1}\IEEEauthorrefmark{3}, Bengü Bilgiç Keskin\IEEEauthorrefmark{1}, \\ İbrahim Hökelek\IEEEauthorrefmark{1}\IEEEauthorrefmark{2}, Ali Görçin\IEEEauthorrefmark{1}\IEEEauthorrefmark{2}, Halim Yanikomeroglu\IEEEauthorrefmark{4}}

\IEEEauthorblockA{\IEEEauthorrefmark{1} \href{https://hisar.bilgem.tubitak.gov.tr/en/} {Communications and Signal Processing Research (HİSAR) Lab., T{\"{U}}B{\.{I}}TAK B{\.{I}}LGEM, Kocaeli, Turkey}}

\IEEEauthorblockA{\IEEEauthorrefmark{2} Department of Electronics and Communication Engineering, Istanbul Technical University, {\.{I}}stanbul, Turkey}
\IEEEauthorblockA{\IEEEauthorrefmark{3} Department of Electrical and Electronics Engineering, Boğaziçi University, {\.{I}}stanbul, Turkey} 

\IEEEauthorblockA{\IEEEauthorrefmark{4} Non-Terrestrial Networks (NTN) Lab, Systems and Computer Engineering, Carleton University, Ottawa, ON, Canada}

Emails: \{omer.karakelle, erhan.karakoca, bengu.bilgic, ibrahim.hokelek\}@tubitak.gov.tr, aligorcin@itu.edu.tr, halim@sce.carleton.ca}

\newacronym{ntn}{NTN}{non-terrestrial networks}
\newacronym{tn}{TN}{terrestrial networks}
\newacronym{nb-iot}{NB-IoT}{narrowband Internet of things}
\newacronym{iot}{IoT}{Internet of things}
\newacronym{leo}{LEO}{low Earth orbit}
\newacronym{ue}{UE}{user equipment}
\newacronym{poc}{PoC}{proof-of-concept}
\newacronym{cfo}{CFO}{carrier frequency offset}

\maketitle

\begin{abstract}
Narrowband Internet of Things (NB-IoT) over non-terrestrial networks (NTN) is a key enabler for massive Internet of Things (IoT) in 6G, but in low Earth orbit (LEO) scenarios, large and time-varying Doppler shifts generate carrier frequency offset (CFO) beyond the correction range of standard user equipment (UE), making initial downlink synchronization a major bottleneck. This paper analyzes Doppler characteristics in realistic NB-IoT LEO scenarios, reviews Doppler mitigation strategies, and proposes a standard-compliant, low-overhead search-space optimization method for downlink acquisition. Results under realistic LEO conditions with real-time measurements show reduced acquisition overhead while maintaining synchronization reliability, supporting NB-IoT adaptation to 6G NTN deployment.

\end{abstract}
\section{Introduction}
The global proliferation of \ac{iot} applications has increased the demand for ubiquitous, reliable, and energy-efficient connectivity solutions. \Ac{tn} typically do not provide adequate coverage in rural, remote, and offshore areas, where infrastructure is limited. \Ac{leo} satellite networks have emerged as an appealing solution to overcome this connectivity bottleneck by enabling \ac{ntn} coverage~\cite{saarnisaari20206g}. To this end, \ac{nb-iot} has gained significance with its cost-effectiveness, low power consumption, and robust performance in challenging environments. Looking ahead to 6G, NTN is expected to become a first-class architectural component for wide-area IoT, resilience, and sustainability; thus, synchronization solutions that are NTN-aware and energy-lean are directly aligned with 6G design targets.

While the use of NTN in wireless networks enhances reliability, LEO satellites typically cause extreme and time-varying dynamic Doppler frequency shift, resulting in a \ac{cfo} that far exceeds the correction range of the \ac{cfo} compensation algorithms implemented by the \ac{nb-iot} \acrfull{ue}. 
While Doppler pre-compensation at the transmitter can mitigate these shifts, its practical implementation is often limited by the UE's localization accuracy and processing constraints. Consequently, there is a growing need for robust receiver-side compensation methods. Recent literature \cite{10227793, 10572045, 10621638, 10892320} has introduced various receiver-side techniques to bridge this gap; however, these designs predominantly target the 5G NR waveform. 
They often do not address NB-IoT specifics, where ultra-narrow bandwidth, repetition mechanisms, and stringent energy budgets fundamentally alter the synchronization trade space.

Crucially, NB-IoT’s initial design and standardization were shaped by terrestrial assumptions, with NTN support was introduced later. This has created a critical research gap where 6G-oriented, NTN-native synchronization strategies are needed. To the best of our knowledge, initial synchronization procedures for \ac{ntn}-\ac{iot} networks mainly focus on uplink transmission \cite{9083854, 9748291}. \textcolor{black}{However, achieving low-complexity initial downlink synchronization is equally as important as uplink synchronization in NB-IoT systems. The information acquired during the initial downlink synchronization process is used for subsequent uplink synchronization procedures. Additionally, the UE utilizes the estimated frequency offset obtained from downlink synchronization to precompensate its carrier frequency before uplink transmission \cite{9590534, 333}. Therefore, the accuracy of downlink synchronization directly dictates the residual frequency offset observed at the gNB.} 

\textcolor{black}{Motivated by these limitations, this work rethinks NB-IoT downlink synchronization for LEO-NTN by introducing a standard-compliant, low-overhead search-space optimization method for initial acquisition.} The proposed approach explicitly targets the overhead–reliability trade-off: it reduces synchronization time while maintaining robust detection and estimation under \emph{time-varying} Doppler.

We develop a channel emulator that captures LEO-specific propagation, including dynamic Doppler and path loss, and integrate it with a 3GPP-compliant end-to-end NB-IoT system for measurement-based evaluation. Receiver implementation respects NB-IoT constraints and features~\cite{36211}, ensuring practical relevance. Experiments in realistic LEO scenarios demonstrate that the proposed method cuts acquisition overhead while improving synchronization accuracy, highlighting the feasibility of NB-IoT over NTN for real-world IoT deployments~\cite{10355106,9650576}. By addressing NB-IoT’s terrestrial-centric legacy with an NTN-aware solution, this work contributes a concrete step toward 6G-ready synchronization for wide-area IoT. 

The rest of the paper is organized as follows. Section \ref{sec:system_model} presents the system architecture and signal model. Section \ref{sec:compensation_leo} describes Doppler shift mitigation methods for NTN downlink. Section \ref{sec:sim_test_results} gives the measurement and simulation results, and Section \ref{sec:conclusion} concludes the paper.

\begin{figure*}
    \centering
    \includegraphics[width=0.85\linewidth]{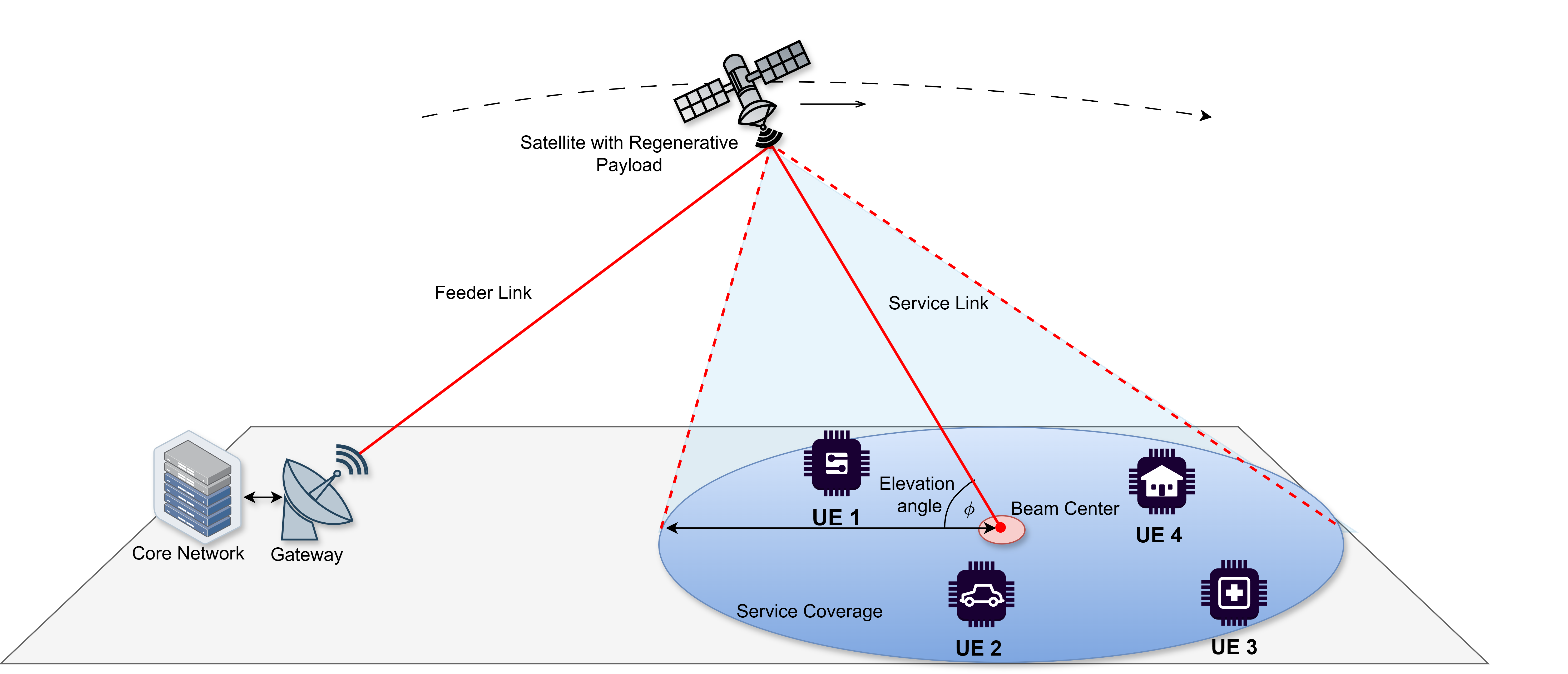}
    \caption{The system model of a satellite IoT network.}
    \label{fig:systemmodel1}
\end{figure*}

\section{System Architecture and Signal Model}
\label{sec:system_model}
In this section, the system architecture and signal model are described for LEO NTN NB-IoT downlink signals employing the Orthogonal Frequency Division Multiplexing (OFDM) waveform.
Fig. 1 shows the considered NB-IoT over LEO-NTN architecture, consisting of a core network, a gateway, a LEO satellite, and multiple NB-IoT UEs within the service area. The gateway communicates with the satellite over the feeder link, while the satellite serves the UEs through the service link. Since this work focuses on NB-IoT downlink synchronization, the main interest is the satellite-to-UE downlink path, specifically the service link. Due to different UE locations within the beam coverage, synchronization conditions vary across users. Based on this architecture, the paper investigates a novel overhead reduction method for downlink synchronization under realistic LEO-NTN conditions.

\subsection{Signal Model}
The OFDM symbol can be defined in discrete time as 

\begin{equation}
s_m[n]=\frac{1}{\sqrt{N}} \sum_{k=0}^{N-1} X_m[k] \exp \left(j 2 \pi \frac{k}{N} n\right),
\label{eq:ofdm_symbols}
\end{equation}

where $X_{m}[k]$ is the data symbol transmitted on the $k^{th}$ subcarrier during the $m^{th}$ OFDM symbol. $N$ is the total number of subcarriers, and ${1/\sqrt{N}}$ is the normalization factor. 
In the NB-IoT downlink, physical channels and signals such as narrowband primary synchronization signal (NPSS), narrowband secondary synchronization signal (NSSS), narrowband physical broadcast channel (NPBCH), and narrowband physical downlink shared
channel (NPDSCH) are mapped onto the OFDM symbols in accordance with the 3GPP specifications. A cyclic prefix (CP) is prepended to each OFDM symbol using the CP size of $N_{CP}$. These CP-OFDM symbols constitute the NB-IoT waveform. The resulting waveform is transmitted using the transmit power of $P_T$. 
\par The LEO channel is modeled by a time-varying channel impulse response (CIR)

\begin{equation}
h(\tau, t)=\sqrt{\frac{1}{PL_{}}} \sum_{\ell=0}^{L-1} \alpha_{\ell} e^{j (2\pi f_Dt+ \theta_{\ell})} \delta\left(\tau-\tau_{\ell}\right), 
\label{eq:channel_model}
\end{equation}

where $PL = PL_{\textsf{fs}} \cdot  PL_{\textsf{a}}$, $PL_{\textsf{fs}}$ is the free space path loss, $PL_{\textsf{a}}$ accounts for the attenuation due to atmospheric gasses \cite{6193}, $L$ is the number of propagation paths, $\alpha_{\ell}$ is the channel gain,
$\theta_{\ell}$ is the phase, $\tau_{\ell}$ is the delay, all corresponding to the $\ell^{\text{th}}$ path, and the Doppler shift $f_D$ given by 

\begin{figure}[ht!]
\centering
\includegraphics[width=0.95\linewidth]{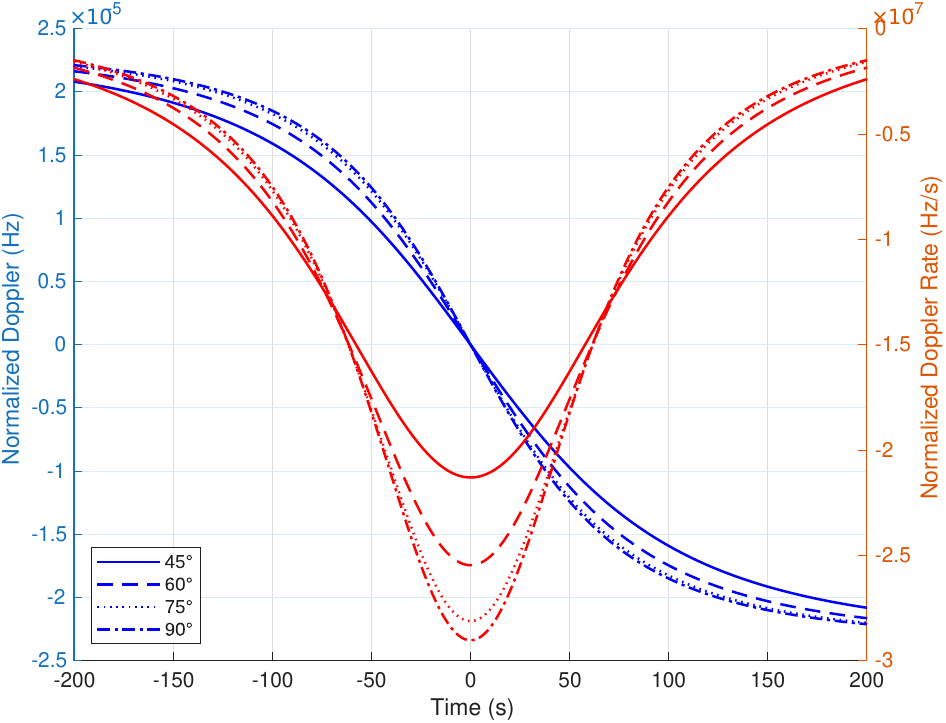}
\caption{Normalized Doppler Shift and Doppler Rate values for different maximum elevation angles.}
\label{fig:doppler}
\end{figure}

\begin{equation}
f_D\, = \frac{ f_c}{c}\frac{\mathbf{V_r} \cdot \mathbf{U_r}}{||\mathbf{U_r}||},
\end{equation}
where $\mathbf{V_r}\,$ is the relative velocity between the UE and the satellite, $\mathbf{U_r}\,$ is the relative position vector directed from the satellite to the UE, $f_c$ is the carrier frequency, and $c$ is the speed of light. According to the ITU recommendation P.619-3 \cite{6193}, $PL_{\textsf{fs}}$ can be expressed in dB as 
\begin{equation}
PL_{f s,dB}=92.45+20 \log _{10}(f_c  d),
\end{equation}
where $d$ denotes the distance in kilometers between the UE and the satellite. 

The UE samples the signal at a rate of $f_s$. The received signal at the UE associated with the $m^{\text{th}}$ CP-OFDM symbol can be represented in discrete-time as
\begin{equation}
r_m\left[n^{\prime}\right]= h_{m}\left[n^{\prime}\right] \circledast \tilde{s}_{m}\left[n^{\prime}\right]+w_m\left[n^{\prime}\right],
\label{eq:discrete_received_signal}
\end{equation}
where $ \quad n^{\prime}=0, \ldots, N, \ldots, N + N_{CP} - 1$, $h_{m}[n']$ is the discrete-time representation of the channel during the $m^{\text{th}}$ symbol, $\tilde{s}_{m}[n']$ and $w_m[n']$ are the $m^{th}$ symbol and the discrete-time AWGN, respectively.

\section{Doppler Estimation and Compensation in NTN LEO NB-IoT}
\label{sec:compensation_leo}

\textcolor{black}{CFO estimation is typically carried out in two main stages: integer and fractional CFO estimation. This study focuses on the integer CFO estimation stage, while the fractional CFO estimation remains outside the scope of this work. To mitigate Doppler shifts caused by the high mobility of LEO satellites, integer CFO estimation is commonly performed through an exhaustive search (ES) over a set of candidate frequencies. This section describes this ES method and proposes a novel search space tailored for communication systems utilizing LEO satellites, as an alternative to the uniform search space commonly adopted in many integer CFO estimation techniques.}

\subsection{CFO Estimation with Exhaustive Search}
An estimate of the Doppler shift, denoted by $\hat{f}_{D}$, can be obtained by analyzing the phase rotation between consecutive pilot symbols. Once estimated, the correction can be applied to the received signal. The frequency corrected received signal can be represented as
\begin{equation}
    r_{\textsf{corr}}[n] = r[n] e^{-j2\pi \frac {\hat{f}_{{D}}}{\Delta f}n/N },
\end{equation}
where $r[n]$ is the overall received signal. The frequency offset estimation is based on a set of candidate frequency offset estimates as

\begin{equation}
C\left(\theta, f_{D,c}\right)=\sum_{n=0}^{K-1} r[n+\theta]   p^*[n]   e^{-j 2 \pi \frac{f_{D,c}}{\Delta f} n / N},
\label{eq:receiver_srsRAN_correlation}
\end{equation}
where $p[n]$ is the known synchronization signal (i.e., second slot of NPSS), $K$ is the size of $p[n]$ and $f_{D,c}$ is a frequency offset hypothesis. 
The estimated integer CFO is obtained as the candidate maximizing the correlation metric
\begin{equation}
\begin{split}
\left(\hat{\theta}, \hat{f}_{D,c}\right)&=\underset{\theta \in \pmb{\theta}, f_{D,c} \in \mathbf{f_{D,c}}}{\arg \max }\left(\left|C\left(\theta, f_{D,c}\right)\right|\right),\\
&s.t. \, \, |C(\theta, f_{D,c})| > \beta
\end{split}
\label{eq:cfo_sweep}
\end{equation}
where $\beta$ represents an adaptive threshold and $\mathbf{f_{D,c}} = [f_{D,c,1}, f_{D,c,2}, ..., f_{D,c,N_c}]$ denotes the integer frequency offset search space \textcolor{black} which is defined as
\begin{equation}
    \mathbf{f_{D,c}} = \left\{ \Delta\gamma \hspace{0.07cm}  m \mid m \in \mathbb{Z},\; -\frac{N_c}{2} < m < \frac{N_c}{2} \right\}, 
    \label{eq:cfo_list}
\end{equation}
where $N_c$ is the size of $\mathbf{f_{D,c}}$ (i.e., the number of candidate frequencies) and $\Delta\gamma$ is a constant step size. If correlation peaks fall below the detection threshold in \eqref{eq:cfo_sweep}, the search space is refined by iteratively shifting all elements toward zero until either a satisfactory correlation peak is detected or the maximum number of trials ($N_{\textsf{max}}$) is reached. In each trial, the step size for every element is tuned using a local spacing between neighboring frequencies.  

Finally, using the CFO and timing offset compensated signal $r_{\textsf{c}}[n]=r[n+\hat{\theta}] e^{-j 2 \pi \frac{\hat{f}_{D,c}}{\Delta f} n /N}$, the fractional frequency offset is estimated as \cite{599949}
\begin{equation}
\hat{f}_{D,f}=\frac{f_s}{2\pi N}\angle\left\{\frac{1}{L}\sum_{n=0}^{L-1} r_c[n]   r_c^*[n+N] \right\},
\label{eq:receiver_srsRAN_cfo_estimation}
\end{equation}
where $L$ is the number of samples used for correlation.
\subsection{Nonuniform Exhaustive Search}
The ES method becomes computationally prohibitive under high Doppler shifts due to the rapidly growing search space. Although increasing the $\Delta\gamma$ interval in \eqref{eq:cfo_list} reduces the list size and time complexity, it also leads to higher CFO estimation errors. A nonuniform search space that accounts for Doppler characteristics can balance this trade-off. 

While the conventional CFO estimation employs ES over uniformly distributed candidate values, real-world Doppler characteristics exhibit nonuniform distribution. As clearly seen in Fig. \ref{fig:doppler}, temporal Doppler profiles demonstrate a higher probability density around the extreme values. 
As the likelihood of experiencing Doppler shifts in these regions is higher, utilizing a fine-grained resolution with a smaller $\Delta\gamma$ value around the extreme values reduces both the CFO estimation error and the time complexity. An approximation to Doppler characteristics is defined as \cite{662636}
\begin{equation}
\begin{aligned}
&\nu(t, \phi_{\textsf{max}},R) = \\
&-\frac{f_c}{c}
\frac{
    w_s R_E R \sin\left(\Delta\psi(t)\right) \cos(\xi(\phi_{\textsf{max}},R))
}{
    \sqrt{R_E^2 + R^2 
    - 2 R_E R \cos\left(\Delta\psi(t)\right) \cos(\xi(\phi_{\textsf{max}},R))}
},
\end{aligned}
\label{eq:doppler_character}
\end{equation}
where $R_E$ and $R$ denote the radius of the Earth and the satellite orbit, respectively, $w_s$ is the angular velocity of the satellite, $\phi_{\textsf{max}}$ is the maximum elevation angle, the term $\Delta\psi(t)=\psi(t)-\psi(t_0)$ is the angular distance between epochs $t$ and $t_0$, and $\xi$ is defined as
\begin{equation}
\xi(\phi,R) = \cos^{-1}\left(\frac{R_E}{R}\cos(\phi)\right)-\phi.
\label{eq:xi}
\end{equation}

The nonuniform \textcolor{black}{search space} is obtained by strategically sampling the expression in \eqref{eq:doppler_character}, under the assumptions $\nu(t_0, \phi_{\textsf{max}},R) = 0$ and $ \psi(t_0) = 0$. The first and last elements of the list correspond to the frequency offset values obtained at the minimum elevation angle $\phi_{\textsf{min}}$, occurring during both satellite acquisition (rising) and departure (setting) phases. By integrating the uniform search space defined in \eqref{eq:cfo_list} with the function given in \eqref{eq:doppler_character}, the optimized nonuniform search space is defined as
\begin{equation}
\begin{aligned}
&\mathbf{\tilde{f}}_{\mathbf{D,c}}= \\
&\left\{\nu\hspace{-0.07cm}\left(\frac{T_d}{N_c}\,m, \phi_{\textsf{max}},R\right)\bigg| \,m \in \mathbb{Z}, -\frac{N_c}{2} <m< \frac{N_c}{2}\right\},
\label{eq:nonuni-es}
\end{aligned}
\end{equation}
\textcolor{black}{
where $T_d$ is satellite observation duration, and is defined as \cite{662636} 
\begin{equation}
T_d\approx \frac{2}{w_s}  \cos^{-1}\left(\frac{\cos(\xi(\phi_{\textsf{min}},R))}{\cos(\xi(\phi_{\textsf{max}},R))}\right).
\end{equation}}

As demonstrated in \eqref{eq:doppler_character} and Fig. \ref{fig:doppler}, the Doppler shift distribution is highly dependent on the actual maximum elevation angle of the satellite pass. In a blind synchronization scenario, the ground receiver lacks prior knowledge of the satellite's trajectory. Therefore, to construct a robust nonuniform search space, we must select a fixed design parameter, denoted as $\phi_{\textsf{max,cand}}$, that minimizes the grid mismatch across all possible satellite passes.

Let $\pmb{\phi^*}$ denote the set of expected actual maximum elevation angles. To evaluate the suitability of a specific candidate $\phi_{\textsf{max,cand}}$, we calculate the aggregate Mean Squared Error (MSE) between the candidate search space ($\mathbf{\tilde{f}}_\mathbf{{D,c}}^{\phi_{\textsf{max,cand}}}$) and the actual Doppler distributions ($\mathbf{\tilde{f}_{D,c}^{{\phi}_{}}}$) induced by every possible true angle $\phi \in \pmb{\phi^*}$. This aggregate error function is defined as:  

\begin{equation}
\begin{aligned}
\Lambda(\phi_{\textsf{max,cand}})
&=\sum_{\phi\in \pmb{\phi^*}}\operatorname{MSE}\left(\mathbf{\tilde{f}}_\mathbf{{D,c}}^{\phi_{\textsf{max,cand}}}, \mathbf{\tilde{f}_{D,c}^{{\phi}_{}}}\right)\\
&=\sum_{\phi\in \pmb{\phi^*}} \frac{1}{N_c}\sum_{i = 1}^{N_c} |\tilde{f}_{D,c,i}^{\phi_{\textsf{max,cand}}}-\tilde{f}_{D,c,i}^{\phi}|^2.
\label{eq:phimax}
\end{aligned}
\end{equation}

The optimal design parameter is then obtained by finding the candidate that minimizes this overall error:
\begin{equation}
\begin{aligned}
\phi_{\textsf{max}}=
\underset{{\phi_{\textsf{max,cand}}\,\in\,\pmb{\phi^*}}}{\arg\min} \space \Lambda (\phi_{\textsf{max,cand}}).
\label{eq:argmin}
\end{aligned}
\end{equation}

\begin{figure}[ht!]
    \centering
\includegraphics[width=1\linewidth]{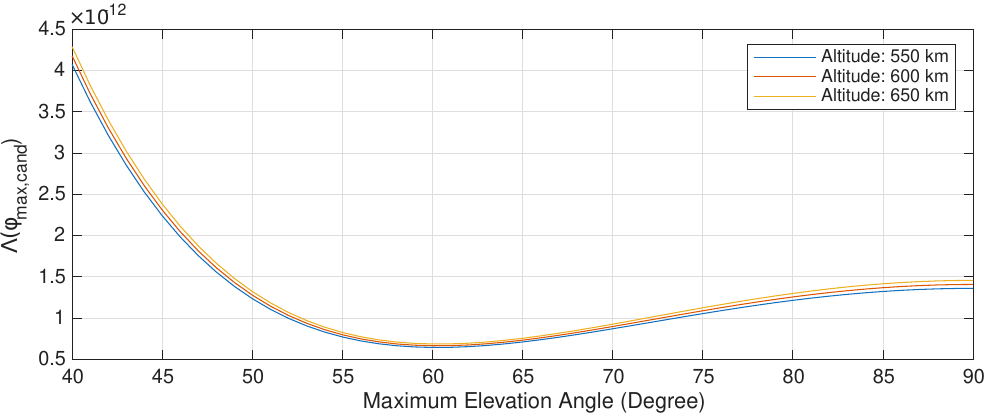}
    \caption{Visualization of the function defined in \eqref{eq:phimax} for $\pmb{\phi^*}=\left\{40^{\circ}, 41^{\circ}, ..., 89^{\circ}, 90^{\circ} \right\}$.}
    \label{fig:mse}
\end{figure} 

\begin{figure*}[ht!]
    \centering
\includegraphics[width=0.8\linewidth]{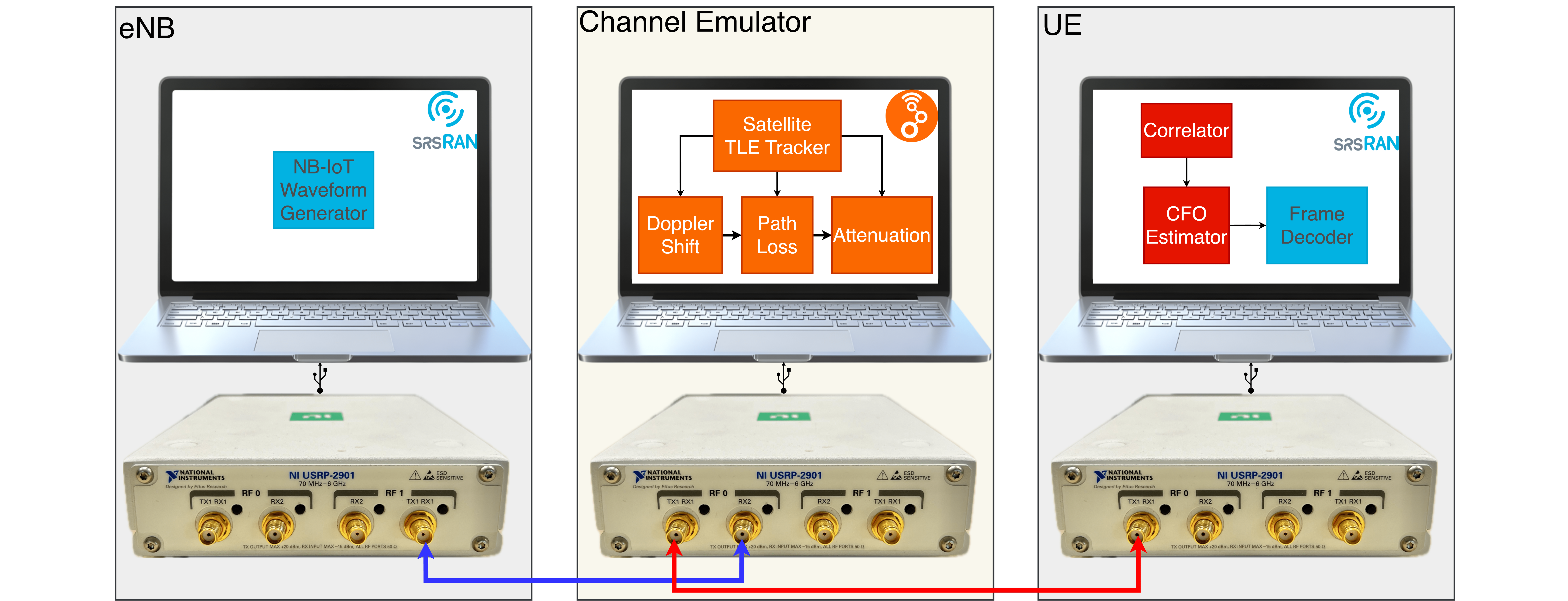}
    \caption{Experimental setup.}
    \label{fig:exp_setup}
\end{figure*}

\section{Simulations and Test Results}
\label{sec:sim_test_results}

This section investigates the downlink \textcolor{black}{synchronization} performance of NB-IoT in NTN environments through a combination of simulations and practical measurements. The scenario depicted in Fig. \ref{fig:systemmodel1} serves as the basis for both measurements and simulations. The measurements are taken throughout a satellite pass for a specific orbital scenario. \textcolor{black}{For the nonuniform search space design, it is obtained that the optimum $\phi_{\textsf{max}}$ is $60^{\circ}$ according to \eqref{eq:argmin} and Fig. \ref{fig:mse}}. The synchronization performance of the ES methods is analyzed based on decoding success rate and the total time elapsed to decode master information block (MIB).

\subsection{Simulation and Measurement Configurations}

\par The CFO estimation success rates of the proposed nonuniform and conventional uniform ES methods are reported by performing the simulation experiments using the parameters in Table I. Both methods employ exhaustive CFO searches over various frequency sweep resolutions, where a larger number of candidate frequencies (i.e., a higher $N_c$ value) implies a fine-grained CFO compensation resolution. 
In the simulations, the UE is assumed to be camped on a known NB-IoT cell, to focus solely on the effectiveness of different Doppler compensation strategies in improving the MIB decoding performance.

Furthermore, NPDSCH is implemented with NTN-compatible enhancements using the srsRAN platform, an open-source radio access network solution. These modifications are specifically applied to the developed NB-IoT UE to support the NTN requirements. The key enhancements include optimizations in cell search and CFO estimation, which address critical challenges for initial synchronization. The simulation environment and experimental setup are designed to replicate a regenerative satellite scenario.

\begin{table}[ht!]
\centering
\caption{Experimental Configurations}
\renewcommand{\arraystretch}{1}
\setlength{\tabcolsep}{5pt}
\label{tab:satellite_trajectory}
\begin{tabular}{lc}
\hline
\multicolumn{1}{l}{\textbf{Parameter}} & \textbf{Value} \\ \hline
\multicolumn{1}{c|}{Bandwidth} & 180 kHz (1 LTE PRB) \\
\multicolumn{1}{c|}{Subcarrier Spacing} & 15 kHz \\
\multicolumn{1}{c|}{Modulation Schemes} & QPSK\\
\multicolumn{1}{c|}{Radio Frame Duration} & 10 ms (10 subframes, 1 ms each) \\
\multicolumn{1}{c|}{Slot Duration} & 0.5 ms \\
\multicolumn{1}{c|}{OFDM Symbols per Slot} & 7 (Normal CP) \\
\multicolumn{1}{c|}{Number of Subcarriers} & 12 \\
\multicolumn{1}{c|}{Cyclic Prefix} & Normal: 4.7 $\mu$s, Extended: 16.7 $\mu$s \\
\multicolumn{1}{c|}{Center Frequency} & 2 GHz\\
\multicolumn{1}{c|}{Sampling Rate} & 1.92 MHz\\
\multicolumn{1}{c|}{Transport Block Size (TBS)} & 24\\
\multicolumn{1}{c|}{Satellite} & STARLINK-1728 \\
\multicolumn{1}{c|}{Observation Duration} & 500 Seconds \\
\multicolumn{1}{c|}{Observation Start Time} & 09.09.2024 07:28\\
\multicolumn{1}{c|}{Observation End Time} & 09.09.2024 07:35 \\
\multicolumn{1}{c|}{Position of the UE} &  $40.789549^{\circ }, 29.451292^{\circ }$  \\
\multicolumn{1}{c|}{Minimum Elevation Angle} & $10^{\circ }$ \\
\multicolumn{1}{c|}{Maximum Elevation Angle} & $86^{\circ }$ \\
\multicolumn{1}{c|}{Maximum Doppler Shift} & 44.3 kHz \\
\hline
\end{tabular}
\end{table}

\subsection{Satellite Trajectory and Ground Station}
The satellite orbit scenario with the selected UE location is shown in Table \ref{tab:satellite_trajectory}, where the STARLINK-1728 satellite is selected as the target orbit. The UE's latitude, longitude, and altitude (LLA) coordinates are set to \( (40.789549^\circ, 29.451292^\circ, 1 \, \text{m}) \). The duration of the experiment is set to 500 seconds, corresponding to the visible pass of the satellite over the reference point.  

\subsection{Test Setup}
Two National Instruments (NI) USRP B210s are used for emulating the RF front ends of the base station (BS) and the UE. Additionally, another NI USRP B210 is utilized as the RF front-end of the channel emulator, which is connected to the UE and the BS with RF cables. All signal processing calculations for emulating the dynamic NTN channel are implemented in C++ within the GNU Radio platform. The channel emulator computations are generated in real-time through the SGP4 algorithm, utilizing the Two-Line Element Set (TLE) orbital parameters of the STARLINK-1728 satellite. The emulator software runs on a high-performance computer system equipped with an Intel Core i9 processor, 64 GB of RAM, and an NVIDIA RTX 4070 8GB GPU, ensuring an accurate setup with powerful processing capabilities. An end-to-end NB-IoT network is established using the open-source srsRAN software stack.

\subsection{Simulation and Measurement Results}

\begin{figure}[ht!]
    \centering
    \includegraphics[width=0.95\linewidth]{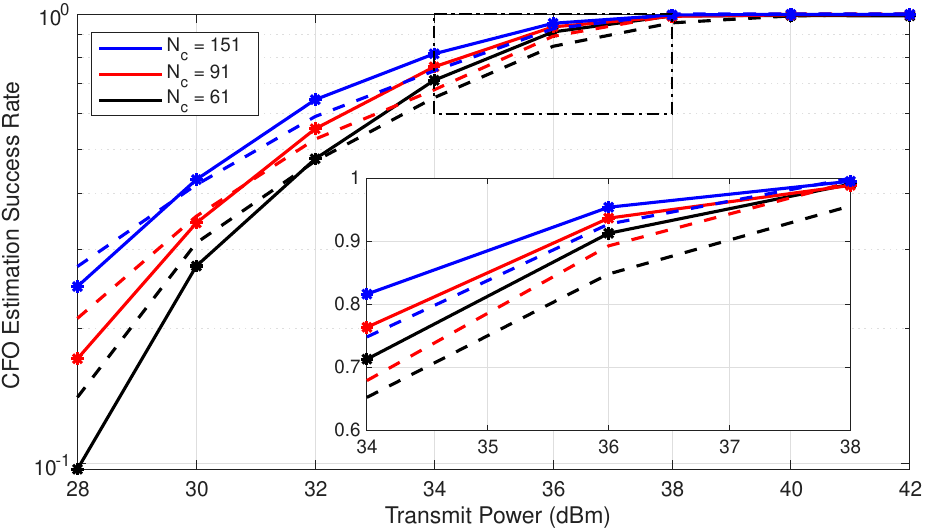}
    \caption{Success rates for different methods (Solid lines: Nonuniform ES, Dashed lines: Uniform ES).}
    \label{fig:matlab_results}
\end{figure} 

\begin{figure}[ht!]
    \centering
    \includegraphics[width=0.95\linewidth]{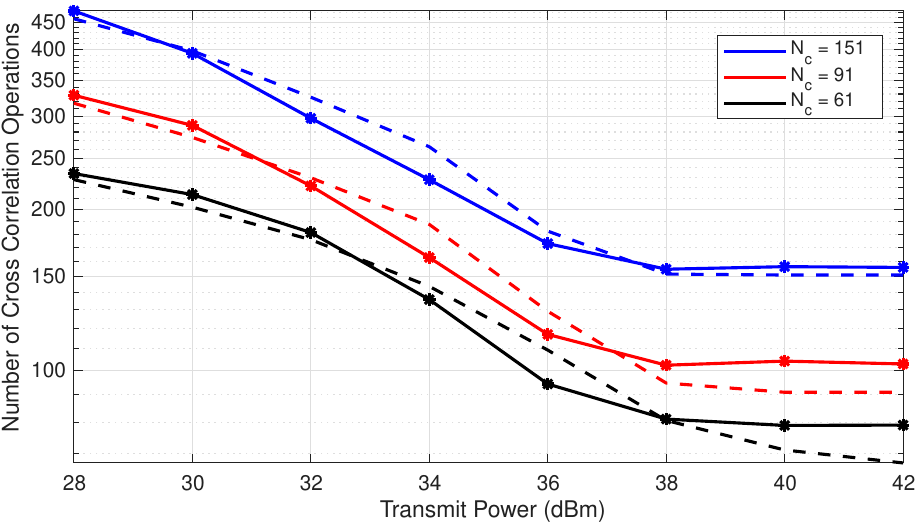}
    \caption{Total number of operations for each method (Solid lines: Nonuniform ES, Dashed lines: Uniform ES).}
    \label{fig:trial_results}
\end{figure} 

\begin{figure}[ht!]
    \centering
    \includegraphics[width=0.95\linewidth]{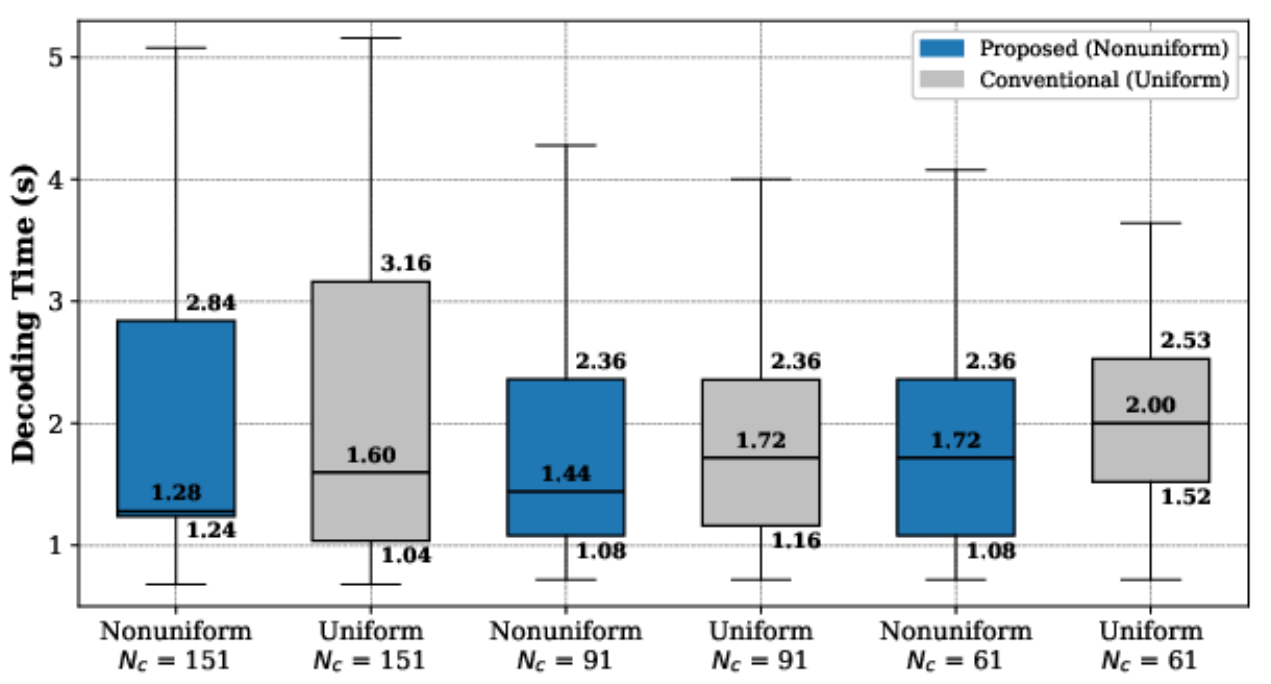}
    \caption{Performance comparison of MIB decoding time of search space methods.}
    \label{fig:meas_res}
\end{figure} 

The CFO estimation success rate results are obtained for $N_{\textsf{max}} = 4$ and different search space sizes as the transmit power is varied from 28 dBm to 42 dBm. Note that each result corresponds to the average of 2000 Monte Carlo simulations. Fig. \ref{fig:matlab_results} shows that the success rate increases when the transmit power increases from 28 dBm to 42 dBm for all experiments. For the same method, a higher success rate is achieved when the number of candidate frequencies (i.e., $N_c$) increases. As clearly seen in Fig. \ref{fig:matlab_results}, the nonuniform ES method (solid lines) outperforms its uniform counterpart (dashed lines). For example, for $N_c = 91$, the nonuniform ES achieves a success rate of 76.3\% at 34 dBm transmit power, which is obtained around 67.9\% for the uniform ES. For $N_c = 151$, the nonuniform ES reaches 81.6\% success rate, while the uniform ES attains only 74.8\%.

Fig. \ref{fig:trial_results} presents a comparative analysis of computational complexity between the proposed nonuniform ES method (solid lines) and the baseline uniform ES approach (dashed lines). The vertical axis represents the number of cross-correlation operations required, while the horizontal axis indicates transmit power levels in dBm. The results demonstrate that the proposed nonuniform method maintains consistently lower computational complexity across most transmit-power levels, achieving reduced operation counts without compromising detection performance.

Fig.~\ref{fig:meas_res} presents the measurement-based performance of the search space methods. Rather than reporting a single average, the boxplots illustrate the distribution of MIB decoding times across 150 independent runs for both the proposed nonuniform and conventional uniform ES methods. Each box highlights the median (indicated by the central bold line), along with the first and third quartiles (Q1 and Q3, representing the lower and upper bounds of the box, respectively).

As depicted in Fig.~\ref{fig:meas_res}, the proposed nonuniform ES method consistently decodes the MIB in less time compared to the uniform ES across the tested configurations. Notably, at $N_c=151$, the proposed method reduces the median decoding time to 1.28 s from 1.60 s compared to the uniform ES, yielding an improvement of 20\%. Furthermore, the nonuniform method maintains lower Q1 and Q3 values in most cases, demonstrating its efficiency and robustness in minimizing synchronization delays. It should also be noted that testbed limitations may introduce minor variations in timing measurements, though the relative performance advantages of the proposed method remain clearly observable across all configurations.

\section{Conclusion} 
\label{sec:conclusion}
This study addresses the initial downlink synchronization challenges in integrated NB-IoT and NTN systems. We proposed a low-complexity, 3GPP-compliant solution and validated its performance through integrated NTN channel emulation and open-source network software. Both simulation and measurement-based evaluations demonstrate that our method enhances robustness and synchronization accuracy. While current NB-IoT standards are terrestrial-oriented, these findings emphasize the necessity for NTN-native enhancements in future 6G releases to further optimize coverage, reliability, and energy efficiency.

\balance
\small
\bibliographystyle{IEEEtran}
\bibliography{main.bib}

\vspace{12pt}
\color{red}
\end{document}